\documentstyle[prc,aps,psfig]{revtex}
\begin{document}
\draft
\title{Color plasma oscillation in strangelets}
\author{Kei Iida}
\address{Department of Physics, University of Tokyo,
7-3-1 Hongo, Bunkyo, Tokyo 113-0033, Japan}
\date{\today}
\maketitle
\begin{abstract}

The dispersion relation and damping rate of longitudinal color
plasmons in finite strange quark matter (strangelets) are evaluated 
in the limits of weak coupling, low temperature, and long wavelength.  
The property of the QCD vacuum surrounding a strangelet makes the 
frequency of the plasmons nearly the same as the color plasma
frequency of bulk matter.  The plasmons are damped by their coupling 
with individual excitations of particle-hole pairs of quarks, of which 
the energy levels are discretized by the boundary.  For strangelets of 
macroscopic size, the lifetime of the plasmons is found to be 
proportional to the size, as in the case of the usual plasma oscillations 
in metal nanoparticles.

\end{abstract}
\pacs{PACS numbers: 12.38.Mh, 25.75.-q}

\section{Introduction}

At extremely high temperature and/or baryon density, quark-gluon
plasmas \cite{B+S} are considered to be energetically favorable as 
compared with hadronic matter.  Such extreme conditions prevent us
from confirming the presence of quark-gluon plasmas, although it is 
expected in interiors of neutron stars and in ultrarelativistic 
heavy-ion collisions.  The intriguing possibility that strange quark 
matter, composed of $u$, $d$, and $s$ quarks, might be the ground
state of the strong interaction was pointed out by Witten \cite{Wit}. 
If this is correct, finite strange quark matter, which is usually 
referred to as strangelets, would be a more stable self-bound system 
than a $^{56}$Fe nucleus.  Then, disruption of neutron stars and 
highly energetic heavy-ion collisions could leave behind many 
strangelets.  This might allow us to detect quark-gluon plasmas in 
the form of strangelets under various experimental situations 
\cite{RUTH,AGS,HIA}.

The ground-state properties of strangelets were described by Farhi 
and Jaffe \cite{FJ} within the MIT bag model; finite-size effects 
such as surface, Coulomb, and shell effects were taken into account. 
In ultrarelativistic heavy-ion collisions, however, the possibly
formed strangelets should be accompanied by thermally excited states.
Among various excitations, we take note of spin- and flavor-symmetric, 
longitudinal color plasmons \cite{BC,Kal} that represent collective 
counteroscillations of different color degrees of freedom.  In a 
strangelet, these plasmons are expected not only to exhibit a
dispersion relation dependent on the color dielectric property of the 
medium, but also to undergo a size-dependent damping as in the case of 
an optically excited metal cluster or nanoparticle.  When the 
nanoparticle is small enough for the electrons to be nearly 
collisionless and for the electric dipole radiation to be of little 
significance but is too large for the shell structure to take effect, 
the surface plasmon excitations are damped in a time proportional to 
the size \cite{KK,Kre}.  Their energy is resonantly absorbed by
individual excitations of particle-hole pairs of the electrons having
discrete energy levels due to the presence of the boundary.
Notice that the quark-gluon plasma in a strangelet is fully
relativistic in contrast to the electron gas in a nanoparticle. 
Consequently, new features should arise from spin-orbit interactions
and from excitations involving antiquarks, both of which may affect 
the lifetimes of the color plasmons.

Longitudinal color plasma oscillations in relativistic and degenerate
bulk matter of $u$, $d$, and $s$ quarks were investigated \cite{BC,Kal} 
by calculating the current-current correlation function within the 
random-phase approximation (RPA) which is valid in the weak-coupling 
limit.  The dispersion law and damping for these oscillations
are essentially the same as those examined by Jancovici \cite{J} for
electromagnetic plasma oscillations in a relativistic and
degenerate electron gas.  In either case, the mode properties are 
characterized by the corresponding plasma frequency and Landau damping
as in a nonrelativistic plasma.  In this paper, we estimate the 
long-wavelength, low-temperature properties of longitudinal color 
oscillation modes in a strangelet of macroscopic size, surrounded by
the QCD vacuum, by incorporating the discrete energy levels of quarks
into the current-current correlation function within the RPA.  The
vanishing color dielectric constant of the vacuum leads to the absence 
of the surface color charge; if present, it would modify the effective
color electric field inside the strangelet according to a Mie-type 
formula.  We thus find that the frequency of the longitudinal color
plasmons in the strangelet is almost the same as that in bulk quark 
matter.  The lifetime of the plasmons is determined by the dipole 
transitions from a quark state in the Fermi sea to an unoccupied one.
It is shown that the quark particle-hole pairs involved damp the
plasmons at a rate inversely proportional to the strangelet size.
Beyond the RPA, however, the plasmons are further damped by collisions
between quarks.  

In Sec.\ II, the longitudinal polarization function is calculated
up to one-loop order, allowing for the discrete quark states inside
a strangelet.  The resulting dispersion relation and damping rate of 
the longitudinal color plasma oscillation are described in Secs.\ III
and IV, respectively.  Concluding remarks are given in Sec.\ V.

\section{Polarization}

We consider a spherical strangelet of radius $R$ and baryon number
$A$, composed of a nearly noninteracting and uniform gas of $u$, $d$,
and $s$ quarks and embedded in the QCD vacuum.  The chemical potential 
$\mu_{i}$ of quarks of flavor $i$, including the rest mass $m_{i}$, is
assumed to be independent of $i$.  Such independence ensures flavor
equilibrium among the quarks, since normally leptons have a vanishingly
small chemical potential in the strangelet \cite{FJ}.  We work in
units in which $\hbar=c=k_{B}=1$.  We set $\mu_{i}\equiv\mu=300$ MeV, 
$m_{u}=m_{d}=0$, and $m_{s}=0$--$300$ MeV.  The temperature $T$ of the 
system is taken to be in the range $0\leq T\ll\mu$.  The 
zero-temperature model for quark matter will thus be adopted below.

Let us now examine the properties of longitudinal color oscillation 
modes in a strangelet by calculating the corresponding proper 
polarization $\Pi_{00}$ in the RPA (one-loop order).  At $T=0$, 
$\Pi_{00}$ can be expressed in terms of the correlation function of
the time component of the quark current ${\cal J}_{\alpha}^{0}$ 
($\alpha=1,\ldots,8$) as 
\begin{equation}
\Pi_{00}(q)=-\frac{i}{V}\int_{V} d{\bf x}\int_{V} d{\bf y}
\int d(x^{0}-y^{0})e^{iq(x-y)}
\left(\frac{g}{2}\right)^{2}
\langle[{\cal J}_{\alpha}^{0}(x),{\cal J}_{\alpha}^{0}(y)]\rangle
\theta(x^{0}-y^{0})\ ,
\end{equation}
with 
\begin{eqnarray*}
{\cal J}_{\alpha}^{0}(x)=\lambda_{ab}^{\alpha}{\bar\psi}_{ai}(x)
\gamma^{0}\psi_{bi}(x)\ ,
\end{eqnarray*}
where $q=(\omega,{\bf q})$ is the four-momentum transfer, 
$V=4\pi R^{3}/3$ is the volume of the strangelet, $g$ is the 
renormalized color coupling constant, $\psi_{ai}$ is the quark field
of color $a$ and flavor $i$, $\lambda_{ab}^{\alpha}$ are the SU(3) 
generators, and $\langle\cdots\rangle$ denotes the expectation value
in the ground state.  The right-hand side of Eq.\ (1) is not summed
over $\alpha$ and is invariant under change in $\alpha$.  In the limit
$V\rightarrow\infty$, this form of $\Pi_{00}(q)$ reduces to the usual
form written for bulk uniform matter.  Hereafter, we assume that the 
momentum transfer $|{\bf q}|$ is far smaller than $R^{-1}$.

The essential part of the present study is to incorporate the discrete
quark levels into the one-loop order calculations of $\Pi_{00}$.  
For such calculations, we first write the quark field in the form of
a mode expansion
\begin{equation}
\psi_{ai}(x^{0},{\bf x})=\sum_{\xi\kappa}[c_{\xi\kappa a i}
\Psi_{\xi\kappa a i}^{(+)}({\bf x})e^{-i\omega_{ki}x^{0}}
+d_{\xi\kappa a i}^{\dagger}
\Psi_{\xi\kappa a i}^{(-)}({\bf x})e^{i\omega_{ki}x^{0}}]\ ,
\end{equation}
where $\Psi_{\xi\kappa a i}^{(+)}$ and $c_{\xi\kappa a i}$ are the 
spinor and the annihilation operator for the quark eigenstate of
energy $\omega_{ki}=\sqrt{k^{2}+m_{i}^{2}}$ characterized by a set of
the quantum numbers $\xi=\{k,j,m\}$, with wave number $k$,  
angular momentum $j$, and its $z$ component $m$, and by the integer 
$\kappa=\pm(j+1/2)$ corresponding to the parity $(-1)^{j\pm1/2}$;
$\Psi_{\xi\kappa a i}^{(-)}$ and $d_{\xi\kappa a i}$ are for the
antiquark eigenstate of energy $\omega_{ki}$.  The spinors 
$\Psi_{\xi\kappa a i}^{(\pm)}$ can be obtained from the Dirac
equation
\begin{equation}
(i\mbox{\boldmath $\gamma\cdot\nabla$}\pm\gamma^{0}\omega_{ki}
-m_{i})\Psi_{\xi\kappa a i}^{(\pm)}=0
\mbox{~~~~~for $r<R$}
\end{equation}
and the boundary condition
\begin{equation}
-i\mbox{\boldmath $\gamma\cdot$}{\hat{\bf r}}
\Psi_{\xi\kappa a i}^{(\pm)}=\Psi_{\xi\kappa a i}^{(\pm)}
\mbox{~~~~~for $r=R$\ ,}
\end{equation}
where $r$ and ${\hat{\bf r}}$ are the length and the unit vector of
the position vector ${\bf x}$, of which the origin is set to be the 
center of the strangelet.  Condition (4) ensures the absence of the 
outgoing probability flux from the boundary.  The analytic solutions 
to Eq.\ (3) under condition (4) were written \cite{MIT,Vas} in the 
context of the MIT bag model; we adopt the solutions satisfying the 
normalization condition $\int_{V}dV 
(\Psi_{\xi\kappa a i}^{(\pm)})^{\dagger}\Psi_{\xi\kappa a i}^{(\pm)}
=1$.  Such solutions read
\begin{equation}
\Psi_{\xi\kappa a i}^{(\pm)}({\bf x})=
\left(\begin{array}{c}
g^{(\pm)}_{k\kappa i}(r) \\ -if^{(\pm)}_{k\kappa i}(r)
\mbox{\boldmath $\sigma\cdot$}{\hat{\bf r}}
\end{array}\right)\phi^{l}_{jm}(\Omega)\chi_{a}\ ,
\end{equation}
with
\begin{equation}
g^{(\pm)}_{k\kappa i}(r)=A_{k\kappa i}^{(\pm)}j_{\kappa}(kr)
+B_{k\kappa i}^{(\pm)}n_{\kappa}(kr)\ ,
\end{equation}
\begin{equation}
f^{(\pm)}_{k\kappa i}(r)=\frac{k}{\pm\omega_{ki}+m_{i}}
[A_{k\kappa i}^{(\pm)}j_{\kappa-1}(kr)+B_{k\kappa i}^{(\pm)}
n_{\kappa-1}(kr)]\ .
\end{equation}
Here $\chi_{a}$ is the color eigenstate, $j_{n}$ and $n_{n}$ are the 
$n$th-order spherical Bessel functions,  
\begin{equation}
\phi^{l=j+1/2}_{jm}(\Omega)=\frac{1}{\sqrt{2(j+1)}}
\left(\begin{array}{c} 
\sqrt{j-m+1}Y^{m-1/2}_{j+1/2}(\Omega) \\ 
-\sqrt{j+m+1}Y^{m+1/2}_{j+1/2}(\Omega)
\end{array}\right)
\mbox{~~~~~for $\kappa=j+1/2$}
\end{equation}
and
\begin{equation}
\phi^{l=j-1/2}_{jm}(\Omega)=\frac{1}{\sqrt{2j}}
\left(\begin{array}{c} 
\sqrt{j+m}Y^{m-1/2}_{j-1/2}(\Omega) \\ 
\sqrt{j-m}Y^{m+1/2}_{j-1/2}(\Omega)
\end{array}\right)
\mbox{~~~~~for $\kappa=-(j+1/2)$}
\end{equation}
are the two-component spherical spinors, and
\begin{equation}
A_{k\kappa i}^{(\pm)}=\left\{\begin{array}{lll}
\displaystyle{\frac{kR}{|j_{\kappa}(kR)|R^{3/2}}}
[\pm2\omega_{ki}R(\kappa\pm\omega_{ki}R)+m_{i}R]^{-1/2}\ ,
         & \mbox{$\kappa=j+1/2$\ ,} \\
\\
         0\ ,
         & \mbox{$\kappa=-(j+1/2)$\ ,}
 \end{array} \right.
\end{equation}
and
\begin{equation}
B_{k\kappa i}^{(\pm)}=\left\{\begin{array}{lll}
         0\ ,
         & \mbox{$\kappa=j+1/2$\ ,} \\
\\
\displaystyle{\frac{kR}{|j_{-\kappa-1}(kR)|R^{3/2}}}
[\pm2\omega_{ki}R(\kappa\pm\omega_{ki}R)+m_{i}R]^{-1/2}\ ,
         & \mbox{$\kappa=-(j+1/2)$\ ,}
 \end{array} \right.
\end{equation}
are the normalization factors.
The energy eigenvalue $E=\pm\omega_{ki}$ may be determined
from the resultant quantization condition \cite{Mar}
\begin{equation}
j_{l=j\pm1/2}(kR)=\mp\frac{k}{E+m_{i}}
j_{l=j\mp1/2}(kR) 
\mbox{~~~~~for $\kappa=\pm\left(j+\frac{1}{2}\right)\ .$} 
\end{equation}
In the 
case of interest here, in which the baryon number $A$ is macroscopic, 
the solutions to Eq.\ (12) approximately read $k=N_{i}\pi/R$ with the 
principal quantum number $N_{i}=1,2,3,\ldots$ .  Here, the asymptotic 
form of $j_{l}(kR)\sim(1/kR)\cos[kR-(l+1)\pi/2]$ for large $kR$ 
has been used, and only quantities of leading order in $kR$ have been 
retained.  The $N_{i}$th solution has a degree of degeneracy, 
$g_{N_{i}}=12(N_{i}+1)(2N_{i}+1)$, stemming from the color, angular
momentum $j=1/2, 3/2,\ldots, 2kR/\pi-3/2$, and parity 
$(-1)^{j\pm1/2}$ \cite{Note}.  We find from this $g_{N_{i}}$ that the 
Fermi wave number of $i$ quarks is
$k_{F,i}\approx(\pi^{4}n_{i}/6)^{1/3}$ with number density
$n_{i}$, which is in turn related to $\mu_{i}$ as 
$\mu_{i}=\omega_{k_{F,i}i}$.

By building the resulting quark field $\psi_{ai}$ into Eq.\ (1) and by
noting $\langle c_{\xi\kappa a i}^{\dagger}c_{\xi\kappa a i}\rangle
=\theta(\mu_{i}-\omega_{ki})$ and $\langle d_{\xi\kappa a i}^{\dagger} 
d_{\xi\kappa a i}\rangle=0$, calculations of $\Pi_{00}$ up to lowest 
order in ${\bf q}$ can be performed straightforwardly.  
The result, 
being of order ${\bf q}^{2}$ and expressed in terms of dipole 
transitions, can be formally divided as 
$\Pi_{00}(q)=\Pi_{00}^{({\rm mat})}(q)+\Pi_{00}^{({\rm vp})}(q)$.
$\Pi_{00}^{({\rm mat})}$ is the matter contribution given by
\begin{eqnarray}
\Pi_{00}^{({\rm mat})}(q)&=&2\left(\frac{g}{2}\right)^{2}\sum_{i}
\sum_{\xi_{1} \xi_{2} \kappa_{1} \kappa_{2}}
\left\{\theta(\mu_{i}-\omega_{k_{1}i})\theta(\omega_{k_{2}i}-\mu_{i})
\frac{}{}\right.
\nonumber \\ & &
\times
\left(\frac{1}{\omega+i\eta+\omega_{k_{1}i}-\omega_{k_{2}i}}
-\frac{1}{\omega+i\eta-\omega_{k_{1}i}+\omega_{k_{2}i}}\right)
{\cal M}^{(qq)}_{i \xi_{1} \xi_{2} \kappa_{1} \kappa_{2}}({\bf q})
+\theta(\mu_{i}-\omega_{k_{1}i})
\nonumber \\ & &
\left.\times
\left(\frac{1}{\omega+i\eta+\omega_{k_{1}i}+\omega_{k_{2}i}}
-\frac{1}{\omega+i\eta-\omega_{k_{1}i}-\omega_{k_{2}i}}\right)
{\cal M}^{({\bar q}q)}_{i \xi_{1} \xi_{2} \kappa_{1} \kappa_{2}}
({\bf q})
\right\}\ ,
\end{eqnarray}
with
\begin{eqnarray}
{\cal M}^{(qq)}_{i \xi_{1} \xi_{2} \kappa_{1} \kappa_{2}}
&=&-\frac{1}{2V}\int_{V}d{\bf x}\int_{V}d{\bf y}
[{\bf q}\mbox{\boldmath $\cdot$}({\bf x}-{\bf y})]^{2}
\Psi_{\xi_{1}\kappa_{1} a_{1} i}^{(+)\dagger}({\bf x})
\Psi_{\xi_{2}\kappa_{2} a_{2} i}^{(+)}({\bf x})
\Psi_{\xi_{2}\kappa_{2} a_{2} i}^{(+)\dagger}({\bf y})
\Psi_{\xi_{1}\kappa_{1} a_{1} i}^{(+)}({\bf y})
\nonumber \\ &=&
\frac{{\bf q}^{2}}{4V}|\langle k_{1},j_{1},\kappa_{1},i,+
|r|k_{2},j_{2},\kappa_{2},i,+\rangle|^{2}
\left[\frac{(j_{1}-m_{1}+1)(j_{1}+m_{1}+1)}{(j_{1}+1)^{2}}
\right.
\nonumber \\ & &
\times
\delta_{j_{1}+1,j_{2}}\delta_{m_{1},m_{2}}\theta(\kappa_{1}\kappa_{2})
+\frac{(j_{1}-m_{1})(j_{1}+m_{1})}{j_{1}^{2}}
\delta_{j_{1}-1,j_{2}}\delta_{m_{1},m_{2}}\theta(\kappa_{1}\kappa_{2})
\nonumber \\ & &
\left.
+\frac{m_{1}^{2}}{j_{1}^{2}(j_{1}+1)^{2}}
\delta_{j_{1},j_{2}}\delta_{m_{1},m_{2}}\theta(-\kappa_{1}\kappa_{2})
\right]\ ,
\end{eqnarray}
\begin{eqnarray}
{\cal M}^{({\bar q}q)}_{i \xi_{1} \xi_{2} \kappa_{1} \kappa_{2}}
&=&-\frac{1}{2V}\int_{V}d{\bf x}\int_{V}d{\bf y}
[{\bf q}\mbox{\boldmath $\cdot$}({\bf x}-{\bf y})]^{2}
\Psi_{\xi_{1}\kappa_{1} a_{1} i}^{(+)\dagger}({\bf x})
\Psi_{\xi_{2}\kappa_{2} a_{2} i}^{(-)}({\bf x})
\Psi_{\xi_{2}\kappa_{2} a_{2} i}^{(-)\dagger}({\bf y})
\Psi_{\xi_{1}\kappa_{1} a_{1} i}^{(+)}({\bf y})
\nonumber \\ &=&
\frac{{\bf q}^{2}}{4V}
|\langle k_{1},j_{1},\kappa_{1},i,+
|r|k_{2},j_{2},\kappa_{2},i,-\rangle|^{2}
\left[\frac{(j_{1}-m_{1}+1)(j_{1}+m_{1}+1)}{(j_{1}+1)^{2}}
\right.
\nonumber \\ & &
\times
\delta_{j_{1}+1,j_{2}}\delta_{m_{1},m_{2}}\theta(\kappa_{1}\kappa_{2})
+\frac{(j_{1}-m_{1})(j_{1}+m_{1})}{j_{1}^{2}}
\delta_{j_{1}-1,j_{2}}\delta_{m_{1},m_{2}}\theta(\kappa_{1}\kappa_{2})
\nonumber \\ & &
\left.
+\frac{m_{1}^{2}}{j_{1}^{2}(j_{1}+1)^{2}}
\delta_{j_{1},j_{2}}\delta_{m_{1},m_{2}}\theta(-\kappa_{1}\kappa_{2})
\right]\ ,
\end{eqnarray}
and $\Pi_{00}^{({\rm vp})}$ is the vacuum contribution given by
\begin{eqnarray}
\Pi_{00}^{({\rm vp})}(q)&=&-2\left(\frac{g}{2}\right)^{2}\sum_{i}
\sum_{\xi_{1} \xi_{2} \kappa_{1} \kappa_{2}}
\left(\frac{1}{\omega+i\eta+\omega_{k_{1}i}+\omega_{k_{2}i}}
-\frac{1}{\omega+i\eta-\omega_{k_{1}i}-\omega_{k_{2}i}}
\right.
\nonumber \\ & &
\left.
-\frac{2}{\omega_{k_{1}i}+\omega_{k_{2}i}}\right)
{\cal M}^{({\bar q}q)}_{i \xi_{1} \xi_{2} \kappa_{1} \kappa_{2}}
({\bf q})\ .
\end{eqnarray}
Here $\eta$ is a positive infinitesimal, and
$\langle k_{1},j_{1},\kappa_{1},i,+
|r|k_{2},j_{2},\kappa_{2},i,\pm\rangle$, 
in which the sign $+$ $(-)$ specifies the quark (antiquark) state, 
denote the radial components of the matrix elements.  These components
vanish when $k_{1}=k_{2}$, or else they are given by 
\begin{eqnarray}
\lefteqn{|\langle k_{1},j_{1},\kappa_{1}=\pm(j_{1}+1/2),i,+|r|
k_{2},j_{2}=j_{1}+1,\kappa_{2}=\pm(j_{1}+3/2),i,+\rangle|^{2}}
\nonumber \\ & &
=\left[\frac{2k_{1}k_{2}}{(k_{1}^{2}-k_{2}^{2})^{2}R}\right]^{2}
\frac{[(\omega_{k_{1}i}+\omega_{k_{2}i})^{2}R^{2}
+(\kappa_{1}+\kappa_{2})(\omega_{k_{1}i}+\omega_{k_{2}i})R]^{2}}
{[2\omega_{k_{1}i}R(\kappa_{1}+\omega_{k_{1}i}R)+m_{i}R]
[2\omega_{k_{2}i}R(\kappa_{2}+\omega_{k_{2}i}R)+m_{i}R]}\ ,
\end{eqnarray}
\begin{eqnarray}
\lefteqn{|\langle k_{1},j_{1},\kappa_{1}=\pm(j_{1}+1/2),i,+|r|
k_{2},j_{2}=j_{1}-1,\kappa_{2}=\pm(j_{1}-1/2),i,+\rangle|^{2}}
\nonumber \\ & &
=\left[\frac{2k_{1}k_{2}}{(k_{1}^{2}-k_{2}^{2})^{2}R}\right]^{2}
\frac{[(\omega_{k_{1}i}+\omega_{k_{2}i})^{2}R^{2}
+(\kappa_{1}+\kappa_{2})(\omega_{k_{1}i}+\omega_{k_{2}i})R]^{2}}
{[2\omega_{k_{1}i}R(\kappa_{1}+\omega_{k_{1}i}R)+m_{i}R]
[2\omega_{k_{2}i}R(\kappa_{2}+\omega_{k_{2}i}R)+m_{i}R]}\ ,
\end{eqnarray}
\begin{eqnarray}
\lefteqn{|\langle k_{1},j_{1},\kappa_{1}=\pm(j_{1}+1/2),i,+|r|
k_{2},j_{2}=j_{1},\kappa_{2}=\mp(j_{1}+1/2),i,+\rangle|^{2}}
\nonumber \\ & &
=\left(\frac{2k_{1}k_{2}R}{k_{1}^{2}-k_{2}^{2}}\right)^{2}
\frac{\left[(\omega_{k_{1}i}-\omega_{k_{2}i})R
+\displaystyle{\frac{2m_{i}}{\omega_{k_{1}i}-\omega_{k_{2}i}}}
+(\kappa_{1}-\kappa_{2})\right]^{2}}
{[2\omega_{k_{1}i}R(\kappa_{1}+\omega_{k_{1}i}R)+m_{i}R]
[2\omega_{k_{2}i}R(\kappa_{2}+\omega_{k_{2}i}R)+m_{i}R]}\ ,
\end{eqnarray}
\begin{eqnarray}
\lefteqn{|\langle k_{1},j_{1},\kappa_{1}=\pm(j_{1}+1/2),i,+|r|
k_{2},j_{2}=j_{1}+1,\kappa_{2}=\pm(j_{1}+3/2),i,-\rangle|^{2}}
\nonumber \\ & &
=\left[\frac{2k_{1}k_{2}}{(k_{1}^{2}-k_{2}^{2})^{2}R}\right]^{2}
\frac{[(\omega_{k_{1}i}-\omega_{k_{2}i})^{2}R^{2}
+(\kappa_{1}+\kappa_{2})(\omega_{k_{1}i}-\omega_{k_{2}i})R]^{2}}
{[2\omega_{k_{1}i}R(\kappa_{1}+\omega_{k_{1}i}R)+m_{i}R]
[2\omega_{k_{2}i}R(-\kappa_{2}+\omega_{k_{2}i}R)+m_{i}R]}\ ,
\end{eqnarray}
\begin{eqnarray}
\lefteqn{|\langle k_{1},j_{1},\kappa_{1}=\pm(j_{1}+1/2),i,+|r|
k_{2},j_{2}=j_{1}-1,\kappa_{2}=\pm(j_{1}-1/2),i,-\rangle|^{2}}
\nonumber \\ & &
=\left[\frac{2k_{1}k_{2}}{(k_{1}^{2}-k_{2}^{2})^{2}R}\right]^{2}
\frac{[(\omega_{k_{1}i}-\omega_{k_{2}i})^{2}R^{2}
+(\kappa_{1}+\kappa_{2})(\omega_{k_{1}i}-\omega_{k_{2}i})R]^{2}}
{[2\omega_{k_{1}i}R(\kappa_{1}+\omega_{k_{1}i}R)+m_{i}R]
[2\omega_{k_{2}i}R(-\kappa_{2}+\omega_{k_{2}i}R)+m_{i}R]}\ ,
\end{eqnarray}
\begin{eqnarray}
\lefteqn{|\langle k_{1},j_{1},\kappa_{1}=\pm(j_{1}+1/2),i,+|r|
k_{2},j_{2}=j_{1},\kappa_{2}=\mp(j_{1}+1/2),i,-\rangle|^{2}}
\nonumber \\ & &
=\left(\frac{2k_{1}k_{2}R}{k_{1}^{2}-k_{2}^{2}}\right)^{2}
\frac{\left[(\omega_{k_{1}i}+\omega_{k_{2}i})R
+\displaystyle{\frac{2m_{i}}{\omega_{k_{1}i}+\omega_{k_{2}i}}}
+(\kappa_{1}-\kappa_{2})\right]^{2}}
{[2\omega_{k_{1}i}R(\kappa_{1}+\omega_{k_{1}i}R)+m_{i}R]
[2\omega_{k_{2}i}R(-\kappa_{2}+\omega_{k_{2}i}R)+m_{i}R]}\ .
\end{eqnarray}
It is convenient to further divide the matter part 
$\Pi_{00}^{(\rm mat)}$ as
$\Pi_{00}^{(\rm mat)}=\Pi_{00}^{(0)}
+\Pi_{00}^{({\rm so})}+\Pi_{00}^{({\bar q}q)}$,
where $\Pi_{00}^{(0)}$, $\Pi_{00}^{({\rm so})}$, and 
$\Pi_{00}^{({\bar q}q)}$ are composed of the matrix elements 
written by Eqs.\ (17) and (18), Eq.\ (19), and Eqs.\ (20)-(22), 
respectively.  We thus find that 
$\Pi_{00}^{(0)}$ $(\Pi_{00}^{({\rm so})})$
arises from excitations of quark particle-hole pairs, accompanied by
the change in the angular momentum of the scattered quark 
$\Delta j\equiv j_{2}-j_{1}=\pm1$ $(\Delta j=0)$; 
$\Pi_{00}^{({\bar q}q)}$ $(\Pi_{00}^{({\rm vp})})$
arises from excitations of quark-antiquark pairs
via scattering of the quark in the filled Fermi (Dirac) sea, 
accompanied by $\Delta j=0,\pm1$, onto the vacant Dirac (Fermi) sea.  
The quantities $\Pi_{00}^{({\rm so})}$, 
$\Pi_{00}^{({\bar q}q)}$, and $\Pi_{00}^{({\rm vp})}$ are 
associated with effects of relativity: the spin-orbit interactions, 
fluctuations of the Fermi sea involving quark-antiquark pairs, and the
second-order vacuum polarization, respectively.  

\section{Dispersion relation}

The dispersion relation of the longitudinal color plasma oscillation 
has been obtained from the one-loop polarization $\Pi_{00}$ 
given by Eqs. (13) and (16).  
By noting the complete analogy with the case of a metal nanoparticle
in a dielectric medium \cite{Kre}, such a relation can be written as  
\begin{equation}
\left[\frac{1}{3}\epsilon_{\rm 0}
-\frac{{\rm Re}\Pi_{00}(q)}{{\bf q}^{2}}\right]
+\frac{2}{3}\epsilon_{\rm vac}=0\ ,
\end{equation}
where $\epsilon_{0}$ is the color dielectric constant of the 
strangelet, having the color carrier part subtracted out, and 
$\epsilon_{\rm vac}$ is that of the surrounding vacuum.  We normally
set $\epsilon_{0}=1$ and $\epsilon_{\rm vac}=0$.  Up to leading order
in $k_{F,i}R$, the summation over $\xi_{1}$, $\xi_{2}$, $\kappa_{1}$, 
and $\kappa_{2}$ in Eqs.\ (13) and (16) may be performed by, if
available, replacing $\sum_{j}\rightarrow\int_{0}^{2kR/\pi}dj$ and
$\sum_{k}\rightarrow(R/\pi)\int_{0}^{k_{F,i}}dk$.  The main
contribution to ${\rm Re}\Pi_{00}$ is 
${\rm Re}\Pi_{00}^{(0)}={\bf q}^{2}\omega_{p}^{2}/3\omega^{2}$, 
where $\omega_{p}=\sqrt{\sum_{i}g^{2}n_{i}/6\mu_{i}}$ is the color 
plasma frequency of bulk quark matter \cite{BC,Kal}.
$|{\rm Re}\Pi_{00}^{({\rm so})}|/{\bf q}^{2}$ 
($|{\rm Re}\Pi_{00}^{({\bar q}q)}+{\rm Re}\Pi_{00}^{({\rm vp})}
|/{\bf q}^{2}$) is suppressed by one power $1/k_{F,i}R$
$(\alpha_{s})$ with respect to $|{\rm Re}\Pi_{00}^{(0)}|/{\bf q}^{2}$
(unity), where $\alpha_{s}=g^{2}/4\pi$ is the QCD fine structure 
constant.  We thus obtain the mode frequency 
$\omega\approx\omega_{p}$.  Its agreement with the bulk value stems 
from absence of the surface color charge due to 
$\epsilon_{\rm vac}=0$.  The values of $\omega_{p}$ calculated as a
function of $\eta_{s}=m_{s}/\mu$ for $\alpha_{s}=0.4$ have been 
exhibited in Fig.\ 1.  We have found that $\omega_{p}$ depends on 
$m_{s}$ only weakly.

\section{Damping rate}

Within the framework of the RPA, the damping of the longitudinal color 
oscillation mode is determined by its resonant energy transfer into 
individual excitations of particle-hole pairs of the quarks whose 
eigenenergies are made discrete by the boundary.  The corresponding 
damping rate may be derived from the polarization $\Pi_{00}$ obtained 
above as \cite{Kal}
\begin{equation}
\Gamma_{\rm RPA}=-\frac{3\omega_{p}}{2{\bf q}^{2}}
{\rm Im}\Pi_{00}(q)|_{\omega=\omega_{p}}\ .
\end{equation}
Here $\Gamma_{\rm RPA}\ll\omega_{p}$ is assumed, and 
the imaginary part of $\Pi_{00}(q)$ is expressed
up to leading order in $k_{F,i}R$ and ${\bf q}$ as
\begin{equation} 
{\rm Im}\Pi_{00}(q)=-\frac{4g^{2}}{\pi^{4}\mu R}{\bf q}^{2}
\sum_{i}\frac{h(\nu,\eta_{i})}{\nu^{3}}\ ,
\end{equation} 
with $\nu=\omega/\mu$, $\eta_{i}=m_{i}/\mu$, and
\begin{eqnarray} 
h(\nu,\eta_{i})&=&\frac{1}{\nu}
\int_{\max(1-\nu,\eta_{i})}^{1}dz
\frac{\sqrt{z^{2}-\eta_{i}^{2}}\sqrt{(z+\nu)^{2}-\eta_{i}^{2}}}
{(2z+\nu)^{2}}
\nonumber \\ & &
\times\left[(z^{2}-\eta_{i}^{2})
-\frac{\pi^{2}\nu(8z^{3}+8\nu z^{2}+4\nu^{2}z+\nu^{3})}
{64z(2z+\nu)}\ln\left|1-\left(\frac{2}{\pi}\right)^{2}
\frac{z^{2}-\eta_{i}^{2}}{z^{2}}\right|
\right.
\nonumber \\ & &
\left.
+\frac{\pi^{2}\nu(8z^{3}+16\nu z^{2}+12\nu^{2}z+3\nu^{3})}
{64(z+\nu)(2z+\nu)}\ln\left|1-\left(\frac{2}{\pi}\right)^{2}
\frac{z^{2}-\eta_{i}^{2}}{(z+\nu)^{2}}\right|
\right.
\nonumber \\ & &
\left.
+\frac{\pi^{2}\nu^{4}}{32z(z+\nu)}
\left(1+\ln\left|1+\frac{2}{\pi}\sqrt{z^{2}-\eta_{i}^{2}}\mu R
\right|\right)
\right]\ .
\end{eqnarray} 
In Fig.\ 2, we have shown 
$\Gamma_{\rm RPA}/\omega_{p}$ as a function of $R$ for various 
values of $\eta_{s}$.  For $\omega\leq\mu+m_{i}$, which is satisfied
at $\omega=\omega_{p}$, ${{\rm Im}\Pi_{00}^{({\rm vp})}}$ cancels out 
${{\rm Im}\Pi_{00}^{({\bar q}q)}}$.  On the other hand, the 
contributions of ${\rm Im}\Pi_{00}^{(0)}$ and 
${{\rm Im}\Pi_{00}^{({\rm so})}}$, the latter being suppressed by one
power $\alpha_{s}$ with respect to the former, lead to the damping of 
which the rate behaves as $\sim c/R$ \cite{Note2}. Such
contributions may be regarded as Landau damping, i.e., a decay of the
collective mode by exciting a single real particle-hole pair.  In this 
case, the surface of the strangelet, which acts as a boundary
condition determining the discrete quark levels rather than as a
scatterer, renders $\Gamma_{\rm RPA}$ nonzero quantum mechanically and
independent of $\alpha_{s}$ in the weak-coupling limit.  As can be 
observed from Fig.\ 2, $\Gamma_{\rm RPA}$ has only a slight dependence
on $m_{s}$.  In the nonrelativistic limit, only the part 
${\rm Im}\Pi_{00}^{(0)}$ remains up to $O(1/k_{F,i}R)$, still yielding 
$\Gamma_{\rm RPA}\propto R^{-1}$.  We remark in passing that for a 
strangelet with radius $R\lesssim3$ fm, which has a baryon number of 
less than $\sim30$, higher-order terms in $(k_{F,i}R)^{-1}$ as ignored 
here may well have consequence to the frequency and damping rate of
the color plasmons.

We next consider the damping of the longitudinal color plasmons due to
collisions \cite{SG} between quarks in the weak-coupling regime.
These collisions, by color exchange, disrupt the self-consistent
fields responsible for the plasmons and add $\Gamma_{\rm col}$ to the
RPA result $\Gamma_{\rm RPA}$.  If one attempts to calculate 
$\Gamma_{\rm col}$ in perturbation theory, one meets an infrared 
divergence that arises from the color magnetic interactions mediated
by transverse gluons.  At least for finite temperature \cite{Note3},
such a divergence cannot be removed even by dynamical screening
\cite{Wel} of the magnetic interactions \cite{BP}.  This indicates
that calculations of $\Gamma_{\rm col}$ up to leading order in 
$\alpha_{s}$ require nonperturbative features of the relativistic 
quark-gluon plasma which may possibly develop a magnetic mass 
$m_{\rm mag}$ and thereby provide an effective cutoff \cite{BI}.  For
a thermal, massless plasma at zero baryon density, where 
$m_{\rm mag}\sim\alpha_{s}T$, one thus obtains 
$\Gamma_{\rm col}\sim\alpha_{s}\ln(1/\alpha_{s})T$ from kinetic theory
\cite{SG}.  This rate arises from the collision integral associated
with one-gluon exchange in the bulk plasma; finite-size corrections of 
$O(1/k_{F,i}R)$ to such a bulk result for $\Gamma_{\rm col}$ should be
of higher order in $\alpha_{s}$ than $\Gamma_{\rm RPA}$, which remains 
in the absence of the collision integral.  At the nonzero strange quark
mass, baryon density, and temperature of interest here, however, the 
values of $m_{\rm mag}$ and hence of $\Gamma_{\rm col}$ are 
essentially unknown.  Therefore, all we can mention is that such a 
collisional damping or the Landau damping due to individual
excitations of quark particle-hole pairs dominates the total damping
rate $\Gamma_{\rm tot}=\Gamma_{\rm RPA}+\Gamma_{\rm col}$, according
to whether the mean free path of the quarks 
$\sim\Gamma_{\rm col}^{-1}$ is small or large compared with $R$.
Here, it is instructive to note that no damping of the plasmons due to 
gluon radiation occurs in the absence of the surface color charge.

\section{Concluding remarks}

We have calculated the dispersion relation and damping
rate of the longitudinal color plasma oscillations in a strangelet in 
the regimes of weak coupling, low temperature, and long wavelength.
For $k_{F,i}R\gg1$, we have found that the frequency of such 
oscillations is almost as large as that of bulk color plasmons and
that the Landau damping yields a lifetime proportional to $R$.
However, the present analysis is accompanied by several uncertainties.
We have confined ourselves to the case of low $|{\bf q}|$ and $T$.  
With increasing $|{\bf q}|$, multipole plasmon excitations should 
develop.  At finite temperature, furthermore, gluonic excitations in
the intermediate state contribute to the proper polarization $\Pi_{00}$; 
within the one-loop approximation, such a contribution is simply added
to the quark-loop contribution without interference \cite{Kal}.  A 
study allowing for those excitations has yet to be done.  Shell
effects in strangelets also remain to be taken into account; such 
effects are predicted to manifest themselves remarkably for 
$A\lesssim100$ \cite{FJ,GJ}.  It is expected from the analogy with the
case of collective excitations in normal nuclei that whether
strangelets are in the vicinity of closed shells or not plays a role
in the properties of the color plasma oscillations.  Third, it is to
be noted that the present analysis is restricted to the weak-coupling
regime $(\alpha_{s}\ll1)$.  In the strongly interacting
nonperturbative regime $(\alpha_{s}\sim1)$, more relevant to a real
system, the damping rate of the color plasmons could be greatly
modified.  These nonperturbative effects should have primary 
significance to the collisional damping.

To conclude, 
we briefly address what type of traces a strangelet, if formed in
heavy-ion collision experiments \cite{QM97} performed at a beam energy
of order 200 GeV per nucleon on a fixed target, could leave via
excitations of the longitudinal color plasmons.  Immediately after 
central nucleus-nucleus collisions at such energies, it is expected
that a quark-gluon plasma would be produced in the resulting matter of
sufficiently high energy density.  Then, the quark-gluon plasma might 
develop into strangelets via pre-freeze-out evaporation of pions,
$K^{+}$, and $K^{0}$ which carry away entropy and antistrangeness from
the system.  This strangeness distillation mechanism, suggested by
Greiner $et$ $al.$ \cite{Gre}, would work as long as strangelets are
stable against strong decay.  Note that the freezeout temperature 
$\sim100$--$200$ MeV \cite{Hei} is comparable with the frequency 
$\omega_{p}$ for realistic values of $\alpha_{s}$ in the range 
$\sim0.4$--$1$.  Longitudinal color plasma oscillations would thus be 
thermally excited in strangelets until the temperature fully drops.
On a time scale of order $\Gamma_{\rm tot}^{-1}$, the excited color 
plasmons would decay into quark particle-hole pairs, leading to
emission of photons and lepton pairs via electromagnetic interactions.
These electromagnetic signals, having energy of order or less than 
$\sim\omega_{p}$, might be observed in the highly energetic heavy-ion 
collisions at CERN's SPS experiments \cite{QM97}.  In order to support
the picture mentioned above, however, estimates allowing for
relaxation processes occurring in such a cooled and decompressed
system, for other mechanisms producing photons and lepton pairs, and
for the uncertainties discussed in the preceding paragraph, are
required.  Elucidation of this problem is beyond the scope of this
paper. 

\section*{Acknowledgments}

The author thanks G. Baym for useful discussion.  He acknowledges the 
hospitality of Aspen Center for Physics, where this work was
initiated.  This work was supported in part by a Grant-in-Aid for 
Scientific Research provided by the Ministry of Education, Science,
and Culture of Japan through Grants Nos.\ 07CE2002 and 199803687.


\begin{figure}
\begin{center}
\leavevmode\psfig{file=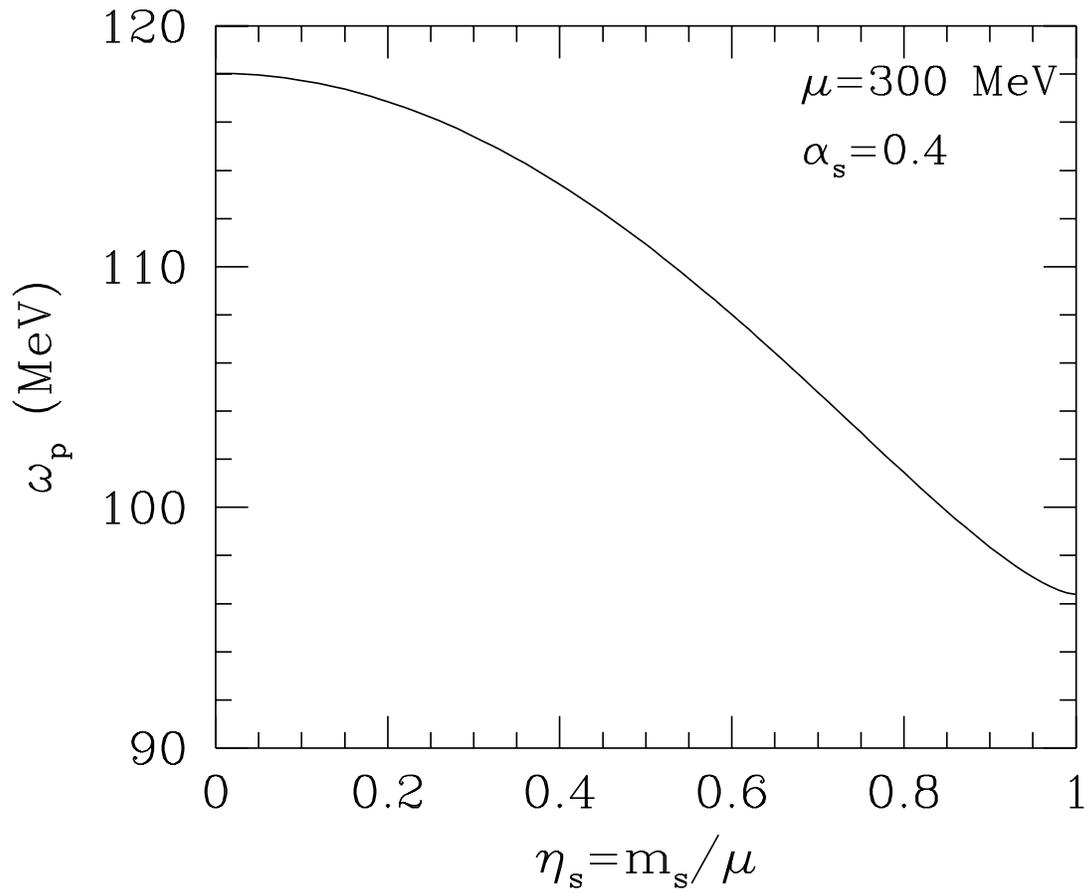,height=15cm}
\caption{Color plasma frequency $\omega_{p}$ as a function of
$\eta_{s}$, calculated for $\mu=300$ MeV and $\alpha_{s}=0.4$.}
\end{center}
\label{fig1}
\end{figure}

\begin{figure}
\begin{center}
\leavevmode\psfig{file=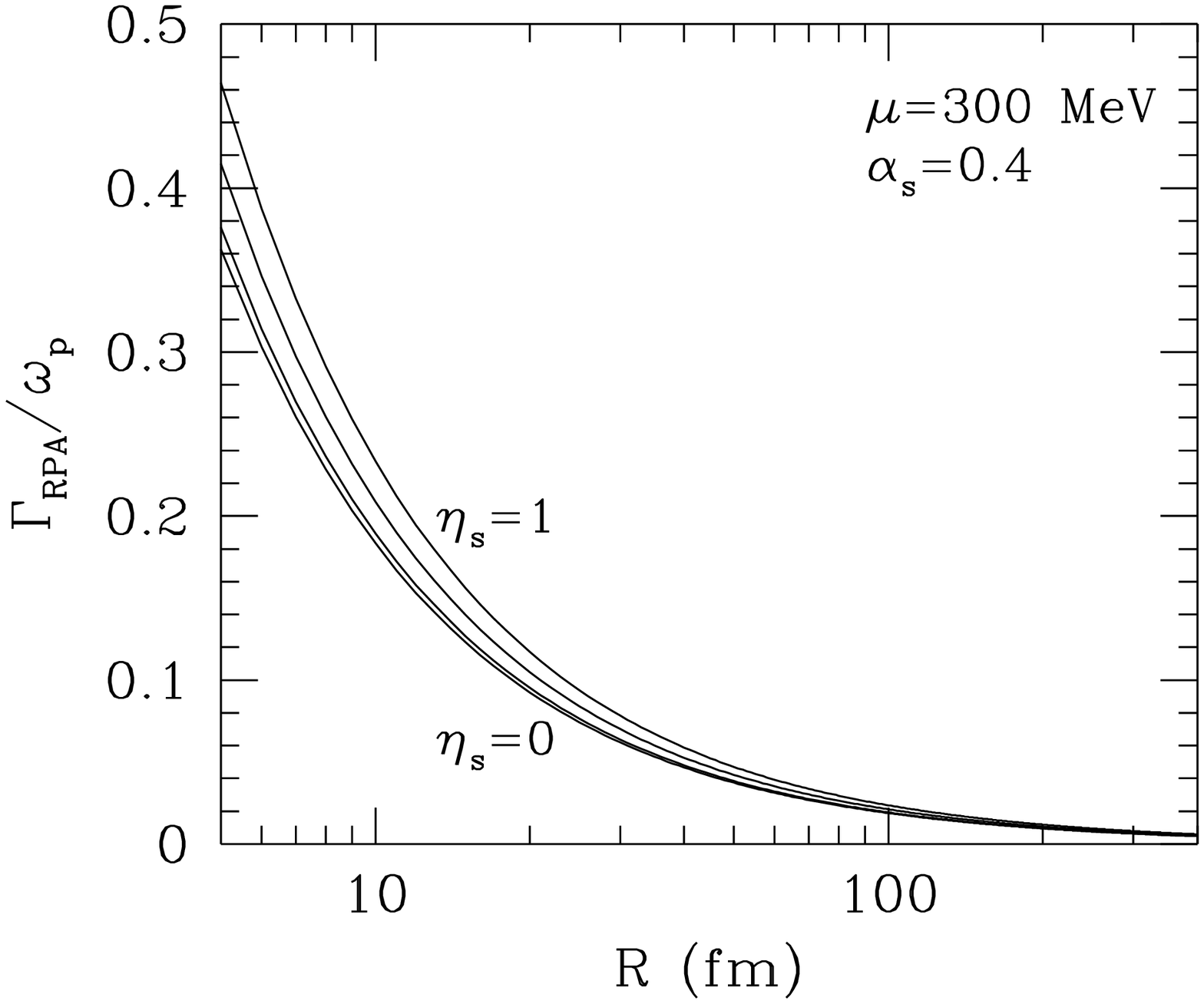,height=15cm}
\caption{Damping rates $\Gamma_{\rm RPA}$ divided by $\omega_{p}$ as a 
function of the strangelet radius $R$, which have been evaluated for 
$\mu=300$ MeV and $\alpha_{s}=0.4$.  The lines from bottom to top are
the results obtained for $\eta_{s}=0,2/3,5/6,1$.} 
\end{center}
\label{fig2}
\end{figure}


\begin{references}
\bibitem{Note}
The rigorous energy spectrum has nondegenerate levels for each 
$(N_{i},j,\kappa)$.  For the large $A$ of interest here, however, it is 
sufficient to note that the eigenstates having $N_{i}$, of which the 
number is identical with $g_{N_{i}}$, lie in the range 
$N_{i-1}\pi/R<k<N_{i+1}\pi/R$.
\end{references}
\end{document}